\documentclass[runningheads]{llncs}

\usepackage[T1]{fontenc}
\usepackage[utf8]{inputenc}
\usepackage{graphicx}
\usepackage{amsmath,amssymb}
\usepackage{bm}
\usepackage{eurosym}

\newcommand{\isrevision}{1}

\ifnum\isrevision=0
    \usepackage[colorinlistoftodos,textsize=scriptsize]{todonotes}
    
    \newcommand{\stkout}[1]{\ifmmode\text{\sout{\ensuremath{#1}}}\else\sout{#1}\fi}
    \newcommand\rev[3]{\textcolor{red}{\begin{scriptsize}{#1}\end{scriptsize}\stkout{#2}}\textcolor{blue}{#3}}
\else
    \usepackage[colorinlistoftodos,textsize=scriptsize, disable]{todonotes}
    \newcommand\rev[3]{\ignorespaces#3\ignorespaces\unskip}
\fi

\begin{document}

\title{From Individual to Shared Ownership: A Coalitional Game Approach to Sustainable Co-investment}

\titlerunning{Coalitional Game Approach to Sustainable Co-investment}

\author{
Yue Yu \and
Andrea Araldo \and
Tijani Chahed \and
Rosario Patan\`e
}

\authorrunning{Y. Yu et al.}

\institute{
SAMOVAR, Télécom SudParis, Institut Polytechnique de Paris, 91120 Palaiseau, France\\
\email{\{yue.yu, andrea.araldo, tijani.chahed, rosario.patane\}@telecom-sudparis.eu}
}

\maketitle

\begin{abstract}
This paper proposes a cooperative game-theoretic framework for sustainable co-investment in shared infrastructure under regulatory incentives. Multiple heterogeneous operators co-invest in a common infrastructure whose production capability evolves over time and is subject to operational variability. A regulator supports the deployment through incentive mechanisms designed to align individual economic investment objectives with the coalitional one.
We formulate the co-investment problem as a transferable-utility (TU) coalitional game in which the value generated by cooperation depends on heterogeneous operational profiles, dynamic resource availability, investment costs, and regulatory incentive level. We show that the proposed coalitional game can be reformulated as a linear production game (LPG)\rev{}{ which enables the construction of stable payoff allocations in which no participant has an incentive to deviate from the coalition}{, whose dual prices yield a constructive and stable allocation of the cooperative surplus}.
\rev{}{Eventually}{Finally}, we illustrate the proposed framework through a case study on co-investment among data center operators in shared renewable energy infrastructure, supported by government subsidies promoting renewable energy consumption.
\end{abstract}


\section{Introduction}

\rev{}{Many infrastructures are traditionally accessed through leasing or pay-per-use mechanisms, such as cloud computing services, transportation fleets, industrial equipment, and energy procurement contracts. While these models reduce upfront investment costs and operational complexity, they may become inefficient when demand grows. Co-investment in shared infrastructure offers an alternative by improving resource utilization, reducing costs and investment risk, and enabling larger-scale deployments.
However, shared ownership introduces coordination challenges. Operators are generally heterogeneous, with different demand profiles, operational constraints, and exposure to uncertainty. Although cooperation may generate substantial collective benefits, some subgroups may still have incentives to deviate if the cooperative surplus is not allocated appropriately. These challenges motivate the use of coalitional game theory, which offers a principled basis for analyzing coalition stability and payoff allocation among self-interested agents.
In this paper, we propose a cooperative game-theoretic framework for sustainable co-investment in shared infrastructure, where a regulator acts as an external incentive provider rather than a strategic player. In this design, cooperation can emerge without incentives, while public support steers it toward more sustainable outcomes.
Our main contributions are as follows:
\begin{itemize}
\item We formulate the co-investment problem as a transferable-utility (TU) coalitional game and define a coalition value function capturing the benefits of shared ownership under heterogeneous operational profiles, regulatory incentives, investment costs, and supply-side variability through a time-varying availability factor.
\item We reformulate the coalition optimization problem as a linear production game (LPG), guaranteeing a non-empty core and yielding a constructive dual-price core allocation.
\item We illustrate the proposed framework through a case study on co-investment among data center operators in shared renewable energy infrastructure supported by government incentives.
\end{itemize}}{Energy- and compute-intensive operators are increasingly exposed to long-term procurement contracts whose costs grow with their scale of consumption. Co-investing in dedicated, shared infrastructure, photovoltaic farms, storage banks, and edge computing facilities, offers a structural alternative, lowering per-unit costs and hedging against external price volatility. However, the parties that benefit from such infrastructure are typically \emph{heterogeneous}: their demand profiles peak at different times, their operational scale differs, and the value they extract from each unit of shared resource depends on when it is supplied. This heterogeneity is precisely what makes cooperation worth analysing.

Coalitional game theory~\cite[Ch.~13]{osborneCourseGameTheory2006} provides the formal language for this problem: a value function over coalitions, a stability concept (the core), and allocation rules that distribute the cooperative surplus. Prior work has applied it to storage sharing~\cite{chakrabortySharingStorageSmart2019,hanIncentivizingProsumerCoalitions2019}, to colocation data centers~\cite{guoAggregationBasedColocationDatacenter2021}, and to cooperative investment under uncertainty~\cite{ketelaarsDynamicStabilityCooperative2025,kiedanskiDiscreteStochasticCoalitional2020,sakrCoInvestmentRevenueUncertainty2025a}. What is missing is a unified treatment that (i) couples \emph{supply-side variability} (a time-varying capacity factor on the shared infrastructure) with \emph{demand-side heterogeneity} and \emph{utilization-based regulatory incentives}, and (ii) yields an allocation that is both a stability certificate and an interpretable solution.

We close this gap by formulating co-investment as a TU game whose value function jointly encodes time-of-use procurement prices, a per-unit utilization subsidy, and a time-varying supply availability. We show that this value function admits a compact formulation as a \emph{linear production game (LPG)}~\cite{owenCoreLinearProduction1975,granotGeneralizedLinearProduction1986}. The dual prices of the grand-coalition LP serve a dual role: they \emph{certify} core non-emptiness and \emph{construct} a core allocation in closed form, in which each operator's share is proportional to the correlation of its demand with the periods of highest marginal value.

Our contributions are:
\begin{itemize}
\item  A coalitional value function for co-investment that integrates time-of-use procurement, utilization-based incentives, and supply-side variability, capturing all four economic levers (prices, subsidy, capacity factor, CAPEX) in a single LP-tractable form (Sec.~\ref{sec:model}).
\item  A constructive dual-price allocation that is provably in the core and admits a per-period economic interpretation, derived without enumerating sub-coalitions (Sec.~\ref{sec:anal}, Appendices~\ref{appendix:lpg_formulation}--\ref{appendix:core_allocation_construction}).
\item A quantitative case study on data center co-investment in shared photovoltaic generation, instantiated with publicly available US benchmark parameters, that (a) quantifies the cooperative surplus, (b) computes the dual-price allocation explicitly, and (c) characterises how a utilization-based subsidy reshapes the optimal coalition capacity (Sec.~\ref{sec:case_study}).
\end{itemize}}

\rev{}{}{The remainder of the paper is organized as follows. Section~\ref{sec:rw} positions our work in the literature. Section~\ref{sec:model} introduces the model. Section~\ref{sec:anal} establishes the LPG formulation and the dual-price core allocation. Section~\ref{sec:case_study} instantiates the framework on data center co-investment case study. Section~\ref{sec:discuss_conclus} concludes the paper.}

\section{Related Work}
\label{sec:rw}

\rev{}{Several studies have applied coalitional game theory to energy sharing and cooperative resource management. Storage-sharing and cooperative energy management have been investigated for smart grids, data centers, and prosumer communities, showing that cooperation can improve resource utilization, reduce procurement uncertainty, and alleviate network stress~\cite{chakrabortySharingStorageSmart2019,guoAggregationBasedColocationDatacenter2021,hanIncentivizingProsumerCoalitions2019}. However, these works mainly focus on operational cooperation or storage sharing, whereas our framework studies the transition from individual procurement to shared infrastructure ownership and explicitly incorporates long-term investment decisions.
Cooperative investment and coalition stability under uncertainty have also been studied using cooperative game theory, real options theory, and LPG formulations~\cite{ketelaarsDynamicStabilityCooperative2025,kiedanskiDiscreteStochasticCoalitional2020}. In addition,~\cite{sakrCoInvestmentRevenueUncertainty2025a} investigates co-investment strategies for edge computing infrastructures. In contrast, our framework jointly integrates regulatory incentives and supply-side variability, and exploits the LPG structure not only to certify core non-emptiness, but also to construct an explicit dual-price allocation of the cooperative surplus.
From a methodological perspective,~\cite{granotGeneralizedLinearProduction1986} introduces a generalized linear production model and dual-based core allocation methods, on which our construction builds. More broadly,~\cite{saadGameTheoreticMethodsSmart2012} surveys game-theoretic methods for smart-grid systems but does not address coalition stability or cooperative co-investment in shared infrastructure.}{We organise prior work into three streams and position our contribution against each.

\textbf{Operational cooperation in energy systems.} Chakraborty et al.~\cite{chakrabortySharingStorageSmart2019} share existing storage among consumers and derive closed-form core allocations; Han et al.~\cite{hanIncentivizingProsumerCoalitions2019} form prosumer coalitions with distributed storage and prove balancedness; Guo et al.~\cite{guoAggregationBasedColocationDatacenter2021} aggregate colocation data center workloads to procure electricity in wholesale markets. These works optimise operational decisions over a \emph{given} asset fleet. We instead model the upstream investment decision, with a value function that jointly captures CAPEX, time-of-use prices, and a utilization-based incentive.

\textbf{Cooperative investment under uncertainty.} Ketelaars et al.~\cite{ketelaarsDynamicStabilityCooperative2025} combine cooperative game theory with real options to characterise dynamic stability of investment coalitions. Kiedanski et al.~\cite{kiedanskiDiscreteStochasticCoalitional2020} formulate shared storage investment as a LPG (LPG) with stochastic demand. Sakr et al.~\cite{sakrCoInvestmentRevenueUncertainty2025a} study co-investment in edge computing under revenue uncertainty. Closest to us is Kiedanski et al., which also exploits LPG structure; we differ on two fronts. First, the variability we model is on the \emph{supply} side, a time-varying capacity factor on the shared infrastructure, rather than on demand. Second, we make the LPG dual prices do double duty: they certify core non-emptiness \emph{and} serve as a per-period pricing mechanism whose contribution per operator is explicit and time-resolved.

\textbf{Methodological foundations.} Owen~\cite{owenCoreLinearProduction1975} introduces LPGs and the dual-price core allocation; Granot~\cite{granotGeneralizedLinearProduction1986} generalises the construction to coalition-dependent resource bundles. Our compact value function fits Owen's original setting after the formulation in Sec.~\ref{sec:model}. Saad et al.~\cite{saadGameTheoreticMethodsSmart2012} survey game-theoretic methods for the smart grid but do not address coalitional investment in shared infrastructure.}

\section{Model and Coalition Value}
\label{sec:model}

We consider a setting in which several operators co-invest in shared  infrastructure \rev{}{in order }{} to meet the resource demands of their services.

\subsection{Players, Time Horizon, and Demand}
We consider a TU game~{\cite[Ch.~13]{osborneCourseGameTheory2006}} with player set
$\mathcal N=\{OP_1,\dots,OP_M\}$ and value function
$v:2^{\mathcal N}\to\mathbb R$.
We consider a discrete time horizon
$\mathcal T=\{0,1,\dots,T\}$,
where $T$ denotes the final time period of the operational lifetime of the shared infrastructure. \rev{}{Each element $t \in \mathcal T$ represents a period over which both resource demand and infrastructure supply are defined.}{}
For each operator $i \in \mathcal{N}$ and time period $t \in \mathcal{T}$,
\rev{}{the resource demand of operator $i$ at time $t$ is denoted as $D_{i,t}$. For any coalition $\mathcal S\subseteq\mathcal N$, the aggregated resource demand is $D_{\mathcal{S},t}=\sum_{i\in \mathcal{S}}D_{i,t}.$}{let $D_{i,t}$ denote the resource demand of operator $i$, with aggregated coalition demand $D_{\mathcal S,t}=\sum_{i\in\mathcal S}D_{i,t}$.}
\rev{}{Operators' services generate an operational benefit $B_{i,t}=B(D_{i,t})$ that depends solely on demand and is independent of how resources are supplied, so $B_{\mathcal S,t}=\sum_{i\in\mathcal S}B_{i,t}$.}{}
Future costs and revenues are discounted by $\delta_t\in(0,1]$\rev{}{ to account for the long-term nature of the investment}{}.

\subsection{Net Benefit under Co-investment}
\rev{}{We now define the net benefit of the coalition formed by operators under co-investment.
}{}Suppose coalition $\mathcal S$ installs a shared infrastructure with capacity $C_\mathcal S \ge 0$.
\rev{}{At time $t$, the effective supply equals $\eta_t C_\mathcal S$,
where $\eta_t \in [0,1]$ is a capacity factor representing the
ratio between the actual supply and the installed capacity.
It captures the variability of the supply generated by the shared infrastructure.}{The effective supply at time $t$ is $\eta_t C_\mathcal S$, where $\eta_t\in[0,1]$ is a capacity factor capturing the time-varying availability of the shared infrastructure.}
Resource supply is first used to satisfy demand; any deficit is
procured from an external provider at unit price $p_t^{\mathrm{in}}$,
while any surplus may be used elsewhere at unit benefit
$p_t^{\mathrm{out}}$. Resources directly consumed by operators from
the shared infrastructure benefit from a regulatory incentive
$\alpha$.
The discounted net benefit associated with capacity level $C_\mathcal S$ is
\begin{equation}
\label{eq:pi_coop}
\begin{aligned}
\pi_{\text{coop}}(\mathcal S,C_\mathcal S)
=
\sum_{t\in\mathcal T}\delta_t
\Big[
&-p_t^{\text{in}}\max(0,D_{\mathcal S,t}-\eta_t C_\mathcal S)
+\alpha\,\min(D_{\mathcal S,t},\eta_t C_\mathcal S)\\
&+p_t^{\text{out}}\max(0,\eta_t C_\mathcal{S}-D_{\mathcal S,t})
\Big]
-k C_\mathcal S,
\end{aligned}
\end{equation}
where $k$ denotes the investment cost per unit of installed capacity.
\rev{}{The discount factor $\delta_t$ accounts for the time value of money,
giving less weight to future costs and revenues compared to present ones.
The case of individual investment is naturally included by considering
singleton coalitions $\mathcal S=\{i\}$.}{}
The no-investment benchmark $\pi_{\text{base}}(\mathcal S) = -\sum_{t\in\mathcal T}\delta_t\, p_t^{\mathrm{in}} D_{\mathcal S,t}$ corresponds to $C_\mathcal S = 0$ in~\eqref{eq:pi_coop}.

\subsection{Coalition Value Function}

\rev{}{The coalition value is defined as the maximal surplus generated by
co-investment relative to the no-investment benchmark.

\begin{definition}[Coalition Value Function under Shared Ownership]
The value of coalition $\mathcal S$ is defined as:
\begin{equation}
\label{eq:value_raw}
v(\mathcal S)
=
\max_{C_\mathcal S\ge0}
\Big(\pi_{\text{coop}}(\mathcal S,C_\mathcal S)-\pi_{\text{base}}(\mathcal S)\Big)
\end{equation}
\end{definition}

\begin{lemma}
The value function \eqref{eq:value_raw} admits the compact form
\begin{equation}
\label{eq:value_reduced}
v(\mathcal S)=
\max_{C_\mathcal S\ge 0}
\left\{
\sum_{t\in \mathcal T}\delta_t \gamma_t\,\min(D_{\mathcal S,t},\eta_t C_\mathcal S)
+\beta C_\mathcal S
\right\},
\end{equation}
where
$\gamma_t=p_t^{\mathrm{in}}-p_t^{\mathrm{out}}+\alpha$ and
$\beta=\sum_{t\in \mathcal{T}}\delta_t p_t^{\mathrm{out}}\eta_t-k$.
\end{lemma}

We assume $\gamma_t\ge0$ and $\beta<0$. The first condition means that local utilization is at least as valuable as export, while the second ensures boundedness of the coalition optimization problem.

We introduce an auxiliary variable
$q_{\mathcal S,t}=\min(D_{\mathcal S,t},\eta_t C_\mathcal S)$,
representing the resources effectively utilized by operators through the shared infrastructure.

The value function~\eqref{eq:value_reduced} is monotone and superadditive in $\mathcal S$, supporting cooperation.}{The coalition value is the maximal surplus generated by co-investment relative to the no-investment benchmark,
\begin{equation}
\label{eq:value_raw}
v(\mathcal S) = \max_{C_\mathcal S\ge0} \big(\pi_{\text{coop}}(\mathcal S,C_\mathcal S)-\pi_{\text{base}}(\mathcal S)\big),
\end{equation}
which admits the compact form
\begin{equation}
\label{eq:value_reduced}
v(\mathcal S) = \max_{C_\mathcal S\ge 0} \left\{\sum_{t\in\mathcal T}\delta_t\gamma_t\,\min(D_{\mathcal S,t},\eta_t C_\mathcal S) + \beta C_\mathcal S\right\},
\end{equation}
with $\gamma_t = p_t^{\mathrm{in}} - p_t^{\mathrm{out}} + \alpha$ and $\beta = \sum_{t\in\mathcal T}\delta_t p_t^{\mathrm{out}}\eta_t - k$. We assume $\gamma_t \ge 0$ and $\beta < 0$: the first ensures local utilization is at least as valuable as export, the second ensures boundedness of the coalition optimization problem. We also introduce an auxiliary variable $q_{\mathcal S,t} = \min(D_{\mathcal S,t}, \eta_t C_\mathcal S)$, representing the resources effectively utilized through the shared infrastructure. The value function~\eqref{eq:value_reduced} is monotone and superadditive in $\mathcal S$, supporting cooperation.}

\section{Analysis}
\label{sec:anal}
\rev{}{In this section, we investigate the stability of cooperation among operators. To derive a stable payoff allocation mechanism, we reformulate the game as a LPG. This formulation enables us to establish the non-emptiness of the core and to construct a payoff allocation belonging to the core.}{We analyze the stability of cooperation among operators. Reformulating the game as a LPG guarantees a non-empty core and yields a constructive dual-price allocation belonging to it.}

\subsection{Core Stability}

\rev{}{In coalitional game theory, the core is the set of payoff allocations for which no coalition has
an incentive to deviate.
\begin{definition}[Core~{\cite[Ch.~13]{osborneCourseGameTheory2006}}]
Let $(\mathcal N,v)$ be a transferable-utility coalitional game.
The core of the game is defined as
\[
\mathrm{Core}(v)=
\left\{
x\in\mathbb R^{|\mathcal N|}
\;\middle|\;
\sum_{i\in\mathcal N}x_i = v(\mathcal N)
\;\text{ and }\;
\sum_{i\in \mathcal S}x_i \ge v(\mathcal S),
\ \forall \mathcal S\subseteq\mathcal N
\right\}.
\]
\end{definition}
A non-empty core thus guarantees the existence of stable allocations, which we establish via a LPG formulation.}{The core of a TU coalitional game $(\mathcal N, v)$~\cite[Ch.~13]{osborneCourseGameTheory2006} is the set of payoff vectors $x \in \mathbb R^{|\mathcal N|}$ satisfying $\sum_{i\in\mathcal N} x_i = v(\mathcal N)$ and $\sum_{i\in\mathcal S} x_i \ge v(\mathcal S)$ for every $\mathcal S \subseteq \mathcal N$, that is, the allocations for which no coalition has an incentive to deviate. A non-empty core thus guarantees the existence of stable allocations, which we establish via a LPG formulation.}

\subsection{LPG Formulation}

\rev{}{The problem admits a LPG structure~\cite[pp.~358--359]{owenCoreLinearProduction1975}: players contribute demand vectors, coalitions pool them, and the coalition value is obtained from a common linear program. By Owen's theorem~\cite[Theorem~1, p.~359]{owenCoreLinearProduction1975}, such games have a non-empty core, and optimal dual prices induce a core allocation.}{In a LPG~\cite[pp.~358--359]{owenCoreLinearProduction1975}, the coalition value is the optimum of a linear program with additive player resources and coalition-independent technology.}

\rev{}{Starting from the reduced value function~\eqref{eq:value_reduced},
the coalition optimization problem can then be written as the linear program}{Starting from~\eqref{eq:value_reduced}, the coalition optimization problem reads}
\begin{equation}
\label{eq:primalqp}
 v(\mathcal S)=
\rev{}{\max_{\mathbf{q_{\mathcal S,t}},C_\mathcal S}}{\max_{\{q_{\mathcal S,t}\}_{t\in\mathcal T},\, C_\mathcal S}}
\left\{
\sum_{t\in \mathcal{T}}\delta_t \gamma_t q_{\mathcal S,t}
+\beta C_\mathcal S
\right\}
\end{equation}
subject to
\begin{align}
q_{\mathcal S,t} &\le D_{\mathcal S,t},
&& t\in \mathcal T, \label{eq:c1}\\
q_{\mathcal S,t}-\eta_t C_\mathcal S &\le 0,
&& t\in \mathcal T, \label{eq:c2}\\
q_{\mathcal S,t} &\ge 0,
&& t\in \mathcal T, \label{eq:c3}\\
C_\mathcal S &\ge 0.
\label{eq:c4}
\end{align}

\begin{proposition}
\label{prop:linear-production}
\rev{}{The coalitional game defined by~\eqref{eq:primalqp}--\eqref{eq:c4} can be reformulated as a LPG~\cite[pp.~358--359]{owenCoreLinearProduction1975}.}{The coalitional game defined by~\eqref{eq:primalqp}--\eqref{eq:c4} is a LPG in the sense of Owen~\cite{owenCoreLinearProduction1975}: each player $i\in\mathcal N$ contributes the demand vector $\{D_{i,t}\}_{t\in\mathcal T}$ as resource endowment, and both the technology matrix and the objective coefficients are independent of the coalition.}

The proof, with explicit vectors, is given in Appendix~\ref{appendix:lpg_formulation}.
\end{proposition}

\subsection{Core Non-Emptiness and Dual-Price Allocation}

\begin{proposition}
\label{prop:core_allocation_construction}
\rev{}{The coalitional game defined by~\eqref{eq:primalqp}--\eqref{eq:c4} has a non-empty core. Moreover, a core allocation based on dual prices is given by $x_i^*=\sum_{t\in \mathcal T} u_t^* D_{i,t}$, where $u_t^*$ denotes the optimal dual prices (Lagrange multipliers) associated with constraint~\eqref{eq:c1}.}{The coalitional game defined by~\eqref{eq:primalqp}--\eqref{eq:c4} has a non-empty core. Let $u_t^*\ge 0$ be the optimal dual variables associated with the demand constraints~\eqref{eq:c1} of the grand coalition. The allocation
\(x_i^*=\sum_{t\in \mathcal T} u_t^*\,D_{i,t}, \ \ i\in\mathcal N,\)
lies in the core. The allocation is constructive: it is obtained from a single LP solution on the grand coalition, with no enumeration of sub-coalitions.}

The proof, based on Owen's dual-price core allocation theorem~\cite[Theorem~1]{owenCoreLinearProduction1975} together with strong duality, is given in Appendix~\ref{appendix:core_allocation_construction}.
\end{proposition}

\subsection{Economic Interpretation}

\rev{}{The dual price $u_t^*$ measures the marginal value of one additional unit of demand at time $t$. Thus, the allocation $x_i^*=\sum_{t\in\mathcal T}u_t^*D_{i,t}$ rewards operators according to both the amount and timing of their resource utilization, giving higher value to demand aligned with periods of productive shared infrastructure.}{Each dual price $u_t^*$ is the marginal value of one additional unit of demand at time $t$ at the grand coalition's optimum. Consequently, $x_i^*=\sum_{t\in\mathcal T}u_t^*\,D_{i,t}$ acts as a per-period pricing mechanism: operator $i$'s share is the inner product of its demand profile with the time-resolved value of the shared resource. Operators whose demand is concentrated in periods of high $u_t^*$, typically periods of high procurement cost or of productive shared supply, receive a proportionally larger share. This makes the allocation both economically interpretable and incentive-compatible with the coalition.}

\section{Case Study}
\label{sec:case_study}

\begin{table}[t!]
\centering
\caption{Main simulation parameters}
\label{tab:simulation_parameters}
\begin{tabular}{lc}
\hline
Parameter & Value \\
\hline
Annual discount rate & $7\%$~\cite{boardmanCBA2017} \\
Off-peak electricity purchase price ($0$--$7$, $22$--$24$) & $0.09$ \$/kWh~\cite{eiaSalesRevenue2023} \\
Mid-load electricity purchase price ($7$--$17$) & $0.15$ \$/kWh~\cite{eiaSalesRevenue2023} \\
Peak electricity purchase price ($17$--$22$) & $0.24$ \$/kWh~\cite{eiaSalesRevenue2023} \\
Grid selling price $p_t^{\mathrm{out}}$ & $0.04$ \$/kWh~\cite{eiaAEO2023} \\
Government subsidy $\alpha$ & $0.04$ \$/kWh~\cite{epaIRA2024} \\
Renewable capacity investment cost $k$ & $1400$ \$/kW installed~\cite{nrelATB2024,eiaAEO2023} \\
Variable demand components $(E_{v,1},E_{v,2},E_{v,3})$ & $(900,700,550)$ kWh per time slot \\
Fixed demand components $(E_{f,1},E_{f,2},E_{f,3})$ & $(600,800,700)$ kWh per time slot \\
\hline
\end{tabular}
\end{table}

\rev{}{To illustrate how the proposed framework can be applied in practice, we now consider a renewable-energy co-investment problem involving multiple data center operators.

data center operators form a coalition $\mathcal S$ to jointly determine the photovoltaic generation capacity $C_\mathcal S$ [kW] to install over an operational horizon $\mathcal T$.
Each time period $t \in \mathcal T$ represents a fixed-duration operational slot. All time-indexed demand and supply quantities are expressed as energy over one time slot [kWh]. Under hourly discretization, these quantities numerically coincide with average power values expressed in [kW] over one hour.
The electricity demand of data center $i$ at time $t$ is modeled as
$D_{i,t}=\rho_{i,t}E_{v,i}+E_{f,i},$
where $\rho_{i,t}\in[0,1]$ denotes the normalized workload level, $E_{v,i}$ [kWh per time slot] represents the workload-dependent electricity consumption component, and $E_{f,i}$ [kWh per time slot] denotes the fixed electricity consumption component. Consequently, $D_{i,t}$ represents the electricity demand of data center $i$ during time slot $t$ and is expressed in [kWh].
Renewable energy production is modeled through the availability factor $\eta_t\in[0,1]$, which captures the variability of photovoltaic generation due to weather conditions. The effective renewable electricity generated during time slot $t$ is therefore given by $\eta_t C_\mathcal S$ [kWh per time slot].
The government subsidizes renewable electricity effectively consumed by the coalition through a utilization-based incentive $\alpha$ [\$/kWh]. Any electricity deficit is purchased from the grid at unit price $p_t^{\mathrm{in}}$ [\$/kWh], while excess renewable electricity is sold back to the grid at unit price $p_t^{\mathrm{out}}$ [\$/kWh].}{We instantiate the framework on a setting that exercises all four economic levers it captures (CAPEX, time-of-use procurement, utilization-based incentive, time-varying supply): three medium-scale data center operators co-investing in a shared photovoltaic plant under a US-style federal subsidy. data centers are a natural test-bed for this analysis: their loads are predictable but heterogeneous across services, they are geographically clusterable, and individual facilities are typically too small to justify utility-scale solar CAPEX alone. 

The coalition $\mathcal S$ jointly determines the photovoltaic generation capacity $C_\mathcal S$ [kW] to install. The operational horizon $\mathcal T$ consists of one hour-resolved reference year repeated over a 20-year project lifetime, with $t$ indexing one-hour slots and all time-indexed quantities expressed as energy per slot [kWh]. The electricity demand of operator $i$ at slot $t$ is modelled as
\(D_{i,t} = \rho_{i,t}\, E_{v,i} + E_{f,i} \ \ \text{[kWh per slot]},\)
where $\rho_{i,t} \in [0,1]$ is a normalized workload level, $E_{v,i}$ the workload-dependent component, and $E_{f,i}$ a fixed baseline. The additive decomposition mirrors the IT-versus-cooling split that dominates data center load profiles. Renewable supply at slot $t$ is $\eta_t C_\mathcal S$, with $\eta_t \in [0,1]$ a time-varying availability factor for the photovoltaic plant. Within the coalition, renewable electricity is pooled and used first to satisfy aggregate demand: the coalition receives the utilization-based incentive $\alpha$ [\$/kWh] on the consumed share, procures any residual deficit from the grid at $p_t^{\mathrm{in}}$ [\$/kWh], and resells any surplus at $p_t^{\mathrm{out}}$ [\$/kWh].

The case study is designed to demonstrate three concrete properties of the framework: (i) in the absence of subsidy, cooperation reduces the total installed capacity yet generates strictly higher coalition value by superadditivity; (ii) a utilization-based subsidy reverses the capacity inequality at the coalition level, steering the grand coalition toward larger renewable deployments; (iii) the dual-price allocation distributes the cooperative surplus stably across the three operators, with each operator's share determined by the time-resolved alignment of its demand with photovoltaic availability.}

\subsection{General Setting}

\begin{figure}[t!]
    \centering
    
    \begin{minipage}{0.48\textwidth}        \includegraphics[width=\textwidth]{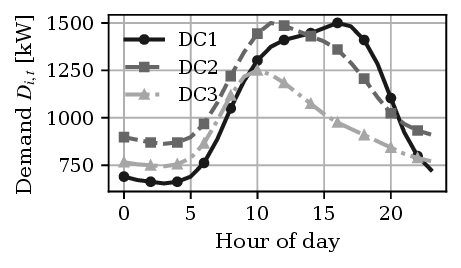}
        \caption{Heterogeneous daily electricity demand profiles of the three data centers}
        \label{fig:workloads}
    \end{minipage}
    \hfill
    \begin{minipage}{0.48\textwidth}
    \centering
    \includegraphics[width=\textwidth]{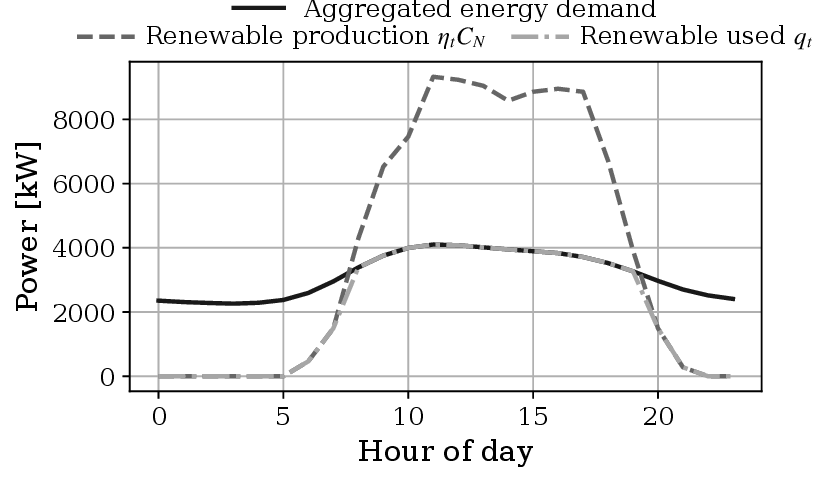}
    \caption{Aggregated electricity demand of the grand coalition and renewable energy supply provided by the shared infrastructure over a representative day}
    \label{fig:energy_supply_demand}
    \end{minipage}
    
\end{figure}

\rev{}{We consider three data centers with heterogeneous daily electricity demand profiles, illustrated in Figure~\ref{fig:workloads}. The duration of each time slot is set to one hour. A reference operational year is considered and repeated over the lifetime of the renewable energy infrastructure.

The shared renewable energy infrastructure is assumed to consist of a photovoltaic solar farm with an operational lifetime of $20$ years. The daily photovoltaic availability profile is illustrated in Figure~\ref{fig:renewable_availability}.
In addition to intra-day variability, the daily profile is further weighted by a seasonal modulation factor.}{We consider three medium-scale data center operators whose daily demand profiles are heterogeneous in both magnitude and time of peak, as illustrated in Figure~\ref{fig:workloads}. The availability factor $\eta_t$ of the shared photovoltaic plant follows the daily solar irradiance and is further weighted by a seasonal modulation factor that captures the variation in solar yield across the year. Figure~\ref{fig:energy_supply_demand} shows the resulting aggregated coalition demand alongside the renewable supply over a representative day, highlighting the temporal mismatch that motivates the capacity sizing problem.}

\begin{figure}[t]
\centering

\begin{minipage}[t]{0.48\textwidth}
\centering
\vspace{0pt}
\includegraphics[width=\textwidth]{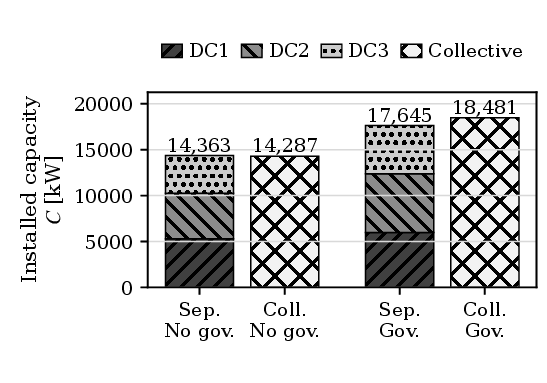}
\caption{Optimal installed renewable energy generation capacities under the four investment scenarios}
\label{fig:scenarios_capacity}
\end{minipage}
\hfill
\begin{minipage}[t]{0.48\textwidth}
\centering
\vspace{0pt}
\includegraphics[width=\textwidth]{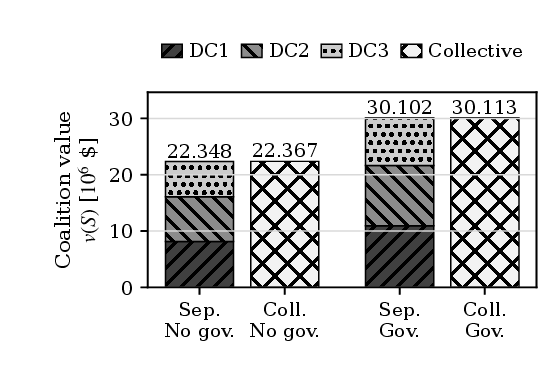}
\caption{Coalition value under the four investment scenarios}
\label{fig:scenarios_value}
\end{minipage}

\end{figure}

\rev{}{A time-dependent electricity purchase price $p_t^{\mathrm{in}}$ is considered. Excess renewable electricity generated by the shared infrastructure is sold back to the grid at a constant unit price $p_t^{\mathrm{out}}$. Future costs and revenues are discounted using an annual discount rate $\delta_t$.}{The grid purchase price $p_t^{\mathrm{in}}$ follows a typical commercial time-of-use schedule with off-peak, mid-load, and peak tiers (Table~\ref{tab:simulation_parameters}); the selling price $p_t^{\mathrm{out}}$ is constant. Discounting is applied at the year level: a slot $t$ belonging to year $y(t) \in \{1,\dots,Y\}$ carries the factor $\delta_t = (1+r)^{-y(t)}$, with $r$ the annual discount rate and $Y$ the project lifetime in years, since the reference year is repeated identically over the lifetime.}

\rev{}{The main simulation parameters are summarized in Table~\ref{tab:simulation_parameters}.}{The parameters in Table~\ref{tab:simulation_parameters} are aligned with publicly available US benchmarks: the time-of-use electricity tariff and the grid selling price reflect EIA retail and AEO data~\cite{eiaSalesRevenue2023,eiaAEO2023}; the $7\%$ discount rate follows the textbook recommendation of~\cite{boardmanCBA2017}, consistent with OMB Circular A-4 guidance for federal public-investment analyses; the PV CAPEX $k$ matches the NREL ATB and EIA AEO benchmarks for utility-scale solar~\cite{nrelATB2024,eiaAEO2023}; the utilization-based incentive $\alpha$ is representative of the IRA Production Tax Credit for solar~\cite{epaIRA2024}, augmented by typical state-level renewable incentives.}

\subsection{Individual Investment versus Co-Investment}

\rev{}{We compare the optimal renewable energy generation capacities under four scenarios combining individual or collective investment with or without government subsidy.
In the individual investment scenarios, each data center independently determines its own optimal renewable energy generation capacity. By contrast, under collective investment, all data centers cooperate and jointly optimize a shared renewable energy infrastructure so as to maximize the total coalition value.
The optimal installed capacities and the associated value achieved under the four scenarios are illustrated in Figure~\ref{fig:scenarios_capacity} and in Figure~\ref{fig:scenarios_value} respectively. 
The results show that, in the absence of government subsidies, cooperation among data center operators leads to a lower total installed capacity compared with independent investment decisions, namely
$C_{\mathcal N}^\ast \sum_{i\in\mathcal N} C_i^\ast.$
This result highlights the efficiency gains generated by shared ownership of renewable energy infrastructure. Because the data centers exhibit heterogeneous workload profiles, renewable electricity can be utilized more efficiently within the coalition than when operators invest independently. Consequently, the same aggregate electricity demand can be supported with a smaller installed renewable energy capacity. A lower installed capacity does not necessarily imply a less favorable outcome. As shown in Figure~\ref{fig:scenarios_value}, cooperation through shared investment generates a higher coalition value than the sum of individual investments because of the superadditivity of the value function.
When government subsidies are introduced, the opposite phenomenon is observed:
$C_{\mathcal N}^\ast>\sum_{i\in\mathcal N} C_i^\ast.$
This result can be explained by the increased economic value of renewable energy consumption induced by the subsidy mechanism. As renewable energy utilization becomes more profitable, the coalition is incentivized to install larger renewable generation capacities in order to maximize the subsidy-related benefits, even if part of the efficiency gains from cooperation is reduced. Furthermore, increasing $\alpha$ increases $C_{\mathcal N}^\ast$, since stronger incentives improve the economic attractiveness of renewable utilization.}{We compare optimal capacities and coalition values across four scenarios crossing individual versus collective investment, with $(\alpha > 0)$ and without $(\alpha = 0)$ the utilization-based incentive (Figures~\ref{fig:scenarios_capacity},~\ref{fig:scenarios_value}).

Without subsidy, cooperation marginally reduces installed capacity below the sum of independent optima: $C_{\mathcal N}^\ast = 14{,}287$ kW versus $\sum_{i\in\mathcal N} C_i^\ast = 14{,}363$ kW. With subsidy the inequality reverses ($18{,}481$ vs $17{,}645$ kW, $+4.7\%$): the coalition installs more capacity to maximise subsidy-eligible consumption, and the effect grows monotonically with $\alpha$. In both regimes the coalition value strictly exceeds the singleton sum (\$22.367\,M vs \$22.348\,M, then \$30.113\,M vs \$30.102\,M), confirming superadditivity; the relative surplus is small because the three workload profiles are only mildly anti-correlated, and stronger temporal complementarity would amplify it without changing the mechanism.

The grand coalition's LP achieves $v(\mathcal N) = \$30.113\,\text{M}$ under subsidy. Solving the dual~\eqref{eq:dual_clean_appendix} gives the optimal multipliers $\{u_t^*\}_{t \in \mathcal T}$ and, via Proposition~\ref{prop:core_allocation_construction}, the constructive core allocation
\((x_1^*,\, x_2^*,\, x_3^*) = (\$10.954\,M,\, \$10.698\,M, \$8.461\,M), \\  \sum_{i\in\mathcal N} x_i^* = v(\mathcal N).\)
Each $x_i^* \ge v(\{i\})$ certifies that no operator has incentive to leave the coalition; the largest share is allocated to the operator whose demand is most concentrated in periods of high $u_t^*$, consistent with the time-resolved pricing interpretation of Sec.~\ref{sec:anal}.}

\section{Conclusion and Perspectives}
\label{sec:discuss_conclus}

\rev{}{This paper proposed a coalitional game framework for sustainable co-investment in shared infrastructure under regulatory incentives. The coalition problem was reformulated as a LPG, guaranteeing a non-empty core and yielding a computable dual-price allocation. The data-center case study showed that cooperation improves investment efficiency and that subsidies can steer the grand coalition toward larger renewable deployments.}{We presented a coalitional game framework for co-investment in shared infrastructure under utilization-based incentives. The coalition value function jointly captures heterogeneous demand, supply-side variability, CAPEX, and time-of-use procurement, and admits a compact linear-production-game formulation whose dual prices play a double role: they certify a non-empty core and construct the core allocation in closed form, where each operator's share equals the inner product of its demand with the time-resolved marginal value of the shared resource.

The data center case study on photovoltaic co-investment, calibrated on US benchmarks, confirmed superadditivity quantitatively, showed that an IRA-style utilization-based subsidy steers the grand coalition toward larger renewable deployments, and assigned the largest share to the operator whose demand best aligned with photovoltaic availability.}

\rev{}{Several extensions deserve further investigation. First, the present model assumes deterministic demand, supply availability, and prices. Future work could introduce stochastic or robust formulations to account for uncertainty in resource demand, infrastructure availability, and external procurement prices. Second, the current model considers a single shared infrastructure capacity decision. Extending the framework to multiple infrastructure types, storage possibilities, or network constraints would enrich the operational layer. Third, the incentive parameter is treated as exogenous. A natural extension would be to model the regulator's decision explicitly, for instance through a bi-level optimization framework in which the regulator chooses the incentive level while operators form coalitions and determine their investment decisions.}{Future directions include stochastic and robust extensions for uncertainty in demand, supply, and prices; multiple infrastructure types, storage, and network constraints; and explicit modelling of the regulator through a bi-level formulation that closes the policy-design loop.}

\section*{Acknowledgments}
This work was carried out in the context of the Celtic Next RAI6Green project, a project funded by the French government as part of the Investments for the future program France 2030.

\bibliographystyle{splncs04}
\bibliography{refs_NETGCOOP}

\newpage

\appendix

\section{LPG Formulation}
\label{appendix:lpg_formulation}
Define the decision vector and the objective coefficient vector
\[
x=(q_{\mathcal S,0},\ldots,q_{\mathcal S,T},C_\mathcal S)^\top\in\mathbb R^{T+2}, \qquad c=(\delta_0\gamma_0,\ldots,\delta_T\gamma_T,\beta)^\top\in\mathbb R^{T+2}.
\]
Constraints~\eqref{eq:c1}--\eqref{eq:c4} can then be written in matrix form as $Ax\le b(\mathcal S)$, $x\ge 0$, where the technology matrix $A\in\mathbb R^{(2T+2)\times(T+2)}$ depends only on $\{\eta_t\}_{t\in\mathcal T}$ and the right-hand side is
\[
b(\mathcal S)=(D_{\mathcal S,0},\ldots,D_{\mathcal S,T},\underbrace{0,\ldots,0}_{T+1})^\top\in\mathbb R^{2T+2}.
\]
Both $A$ and $c$ are independent of $\mathcal S$: only $b(\mathcal S)$ varies with the coalition.
Associating each player $i\in\mathcal N$ with the resource vector
\[
b_i=(D_{i,0},\ldots,D_{i,T},0,\ldots,0)^\top\in\mathbb R^{2T+2},
\]
the additivity $D_{\mathcal S,t}=\sum_{i\in\mathcal S}D_{i,t}$ immediately yields $b(\mathcal S)=\sum_{i\in\mathcal S}b_i$: each coalition's resources are the sum of the resources contributed by its members. Hence~\eqref{eq:primalqp}--\eqref{eq:c4} fits the LPG template of~\cite[pp.~358--359]{owenCoreLinearProduction1975}:
\[
v(\mathcal S)=\max\{c^\top x:\,Ax\le b(\mathcal S),\,x\ge 0\}, \qquad b(\mathcal S)=\sum_{i\in\mathcal S}b_i.
\]

\section{Core Allocation Construction}
\label{appendix:core_allocation_construction}

Since~\eqref{eq:primalqp}--\eqref{eq:c4} has the LPG structure established in Proposition~\ref{prop:linear-production}, Owen's theorem~\cite[Theorem~1, p.~359]{owenCoreLinearProduction1975} implies the game is totally balanced, and the Bondareva--Shapley theorem~\cite[Ch.~13]{osborneCourseGameTheory2006} then yields a non-empty core.

Associating multipliers $u_t\ge 0$ with the demand constraints~\eqref{eq:c1} and $v_t\ge 0$ with the capacity constraints~\eqref{eq:c2}, the dual of~\eqref{eq:primalqp}--\eqref{eq:c4} reads
\begin{equation}
\label{eq:dual_clean_appendix}
\min_{u,\,v\,\ge\, 0}\ \sum_{t\in\mathcal T} u_t D_{\mathcal S,t}
\quad\text{s.t.}\quad u_t+v_t\ge\delta_t\gamma_t\ \forall t\in\mathcal T,\quad \sum_{t\in\mathcal T}\eta_t v_t\le -\beta.
\end{equation}
By strong duality, its optimum equals $v(\mathcal S)$.

By the dual-price core allocation theorem~\cite[pp.~361--362]{owenCoreLinearProduction1975}, an optimal dual solution $(u^*,v^*)$ associated with the grand coalition induces the core allocation
\[
x_i^*=\sum_{t\in\mathcal T} u_t^*\,D_{i,t},\qquad i\in\mathcal N.
\]

The theorem's assumptions are satisfied: the resource vector is additive, $b(\mathcal S)=\sum_{i\in\mathcal S}b_i$ (Appendix~\ref{appendix:lpg_formulation}); $A$ and $c$ are coalition-independent; the grand coalition's primal is feasible ($q_{\mathcal N,t}=C_{\mathcal N}=0$) and bounded above under $\beta<0$, ensuring existence of an optimal dual solution.

\end{document}